# High-Chern-Number and High-Temperature Quantum Hall Effect without Landau Levels

Jun Ge[1†], Yanzhao Liu[1†], Jiaheng Li[2†], Hao Li[3,4†], Tianchuang Luo[1], Yang Wu[4,5], Yong Xu[2,6,7,8*] & Jian Wang[1,8,9,10*]

[1] *International Center for Quantum Materials, School of Physics, Peking University, Beijing 100871, China*

[2] *State Key Laboratory of Low Dimensional Quantum Physics, Department of Physics, Tsinghua University, Beijing 100084, P. R. China*

[3] *School of Materials Science and Engineering, Tsinghua University, Beijing, 100084, P. R. China*

[4] *Tsinghua-Foxconn Nanotechnology Research Center and Department of Physics, Tsinghua University, Beijing 100084, P. R. China*

[5] *Department of Mechanical Engineering, Tsinghua University, Beijing, 100084, P. R. China*

[6] *Frontier Science Center for Quantum Information, Beijing, China*

[7] *RIKEN Center for Emergent Matter Science (CEMS), Wako, Saitama 351-0198, Japan*

[8] *Collaborative Innovation Center of Quantum Matter, Beijing 100871, China*

[9] *CAS Center for Excellence in Topological Quantum Computation, University of Chinese Academy of Sciences, Beijing 100190, China*

[10] *Beijing Academy of Quantum Information Sciences, Beijing 100193, China*

[†]These authors contributed equally to this work.

*Emails: jianwangphysics@pku.edu.cn (J.W.); yongxu@mail.tsinghua.edu.cn (Y.X.)

**ABSTRACT**

The quantum Hall effect (QHE) with quantized Hall resistance of $h/ve^2$ starts the research on topological quantum states and lays the foundation of topology in physics. Afterwards, Haldane proposed the QHE without Landau levels, showing nonzero Chern number $|C|=1$, which has been experimentally observed at relatively low temperatures. For emerging physics and low-power-consumption electronics, the key issues are how to increase the working temperature and realize high Chern numbers ($C>1$). Here, we report the experimental discovery of high-Chern-number QHE ($C=2$) without Landau levels and $C=1$ Chern insulator state displaying nearly quantized Hall resistance plateau above the Néel temperature in $MnBi_2Te_4$ devices. Our observations provide a new perspective on topological matter and open new avenues for exploration of exotic topological quantum states and topological phase transitions at higher temperatures.

**Keywords:** quantum Hall effect without Landau levels, Chern insulator, topological matter, topological quantum states, high Chern number, high temperature

# INTRODUCTION

The quantum Hall effect (QHE) with quantized Hall resistance plateaus of height $h/ve^2$ was firstly observed in two-dimensional (2D) electron systems in 1980 [1]. Here, $h$ is Planck's constant, $v$ is Landau filling factor and $e$ is electron charge. The QHE in the 2D electron systems with high mobility is originated from the formation of Landau levels (LLs) under strong external magnetic field. Subsequently, the exact quantization was explained by Laughlin based on gauge invariance and was later related to a topological invariance of the energy bands, which is characterized by Chern number $C$ [2-5]. Nonzero Chern number distinguishes the QHE systems from vacuum with $C=0$ [2, 3]. The discovery of QHE introduces the concept of topology into condensed matter physics and is extremely important to physical sciences and technologies. However, the rigorous conditions of ultrahigh mobility, ultralow temperature and strong external magnetic field limit the deep exploration and wide applications of QHE.

Theoretical proposals based on the intrinsic band structure of 2D systems open up new opportunities. In 1988, Haldane theoretically proposed a time-reversal symmetry (TRS) breaking 2D condensed-matter lattice model with quantized Hall conductance of $e^2/h$ in the absence of external magnetic field [6]. This indicates that QHE can be realized without the formation of LLs. The QHE induced by spontaneous magnetization in such insulators is called quantum anomalous Hall effect (QAHE), and such insulators are called Chern insulators. Alternative mechanism of realizing QAHE through localization of band electrons was later proposed in 2003 [7]. The emergence of topological insulators (TIs) in which strong spin-orbit coupling (SOC) gives rise to topological band structures provides a new platform for the investigation on QHE without strong external magnetic field. The QAHE with quantized Hall conductance of $e^2/h$ was predicted to occur in magnetic TIs by doping transition metal elements (Cr or V) into time-reversal-invariant TIs $Bi_2Te_3$, $Bi_2Se_3$, and $Sb_2Te_3$ [8]. In 2013, the QAHE with quantized Hall conductance of $e^2/h$ was experimentally observed in thin films of chromium-doped $(Bi,Sb)_2Te_3$ at the temperature down to 30 mK[9].

However, in the above mentioned QHE systems without LLs, only Hall resistance plateau with $C=1$ can be obtained by coupling topological surface states with magnetism. High-Chern-number QHE without LLs has never been observed experimentally. Besides, the requirement of ultralow temperatures limits the study of QHE without LLs. Efforts on high-Chern-number and high-temperature QHE without LLs are still highly desired for exploring emergent physics and low-power-consumption electronics [10].

Here we report the first experimental discovery of the high-Chern-number QHE without LLs above 10 K and $C=1$ QHE without LLs above the Néel temperature ($T_N$) in $MnBi_2Te_4$ devices. We show that when modulated into the insulating regime by a small back gate voltage, the nine-layer and ten-layer $MnBi_2Te_4$ devices can be driven to Chern insulator with $C=2$ at moderate perpendicular magnetic field. Quantized Hall resistance $h/2e^2$ accompanied with vanishing longitudinal resistance at the temperature as high as 13 K is observed in the ten-layer device. When reducing the thickness of the devices down to eight-layer and seven-layer, quantized Hall resistance plateau $h/e^2$ is detected at the temperature much higher than the Néel temperature of the devices. This quantization temperature is the highest record in systems showing QHE without LLs. Our discoveries break new ground in the exploration of topological quantum states and provide a platform for potential applications in related low-consumption electronics.

$MnBi_2Te_4$ is a layered material which can be viewed as a layer of $Bi_2Te_3$ TI intercalated with an additional Mn-Te bilayer [11-20]. This material exhibits ferromagnetic (FM) order within septuple layer (SL) and anti-ferromagnetic (AFM) order between neighboring SLs with an out-of-plane easy axis [11], as displayed in Fig. 1a. By tuning the magnetic structure through thickness or magnetic field, exotic topological states, such as type-I topological Weyl semimetal (WSM) in 3D, Chern insulator in 2D and higher-order topological Möbius insulator, can be realized in $MnBi_2Te_4$ [21, 22]. In this work, the $MnBi_2Te_4$ flakes were mechanically exfoliated from high quality $MnBi_2Te_4$ single crystals. These flakes were then transferred to 300 nm $SiO_2$/Si substrates and the standard e-beam lithography followed by e-beam evaporation was used to fabricate electrodes. The doped Si served as the back gate and a back gate voltage applied between Si and the sample can modulate the sample into insulating regime.

## RESULTS

### High-Chern-number Chern insulator states

Figure 1b shows an optical image of the $MnBi_2Te_4$ device (s6) with Hall bar geometry. Atomic force microscope measurements were carried out to determine the thickness of s6 (Fig. S1f). The line profile reveals a thickness of 13.4 ± 0.4 nm, corresponding to 10-SL. The temperature dependence of longitudinal resistance $R_{xx}$ is shown in Fig. 1c, in which a sharp resistance peak gives the $T_N$ at around 22 K.

To get insight into the evolution of the Chern insulator states in the 10-SL $MnBi_2Te_4$ device s6, we carried out magneto-transport measurements at various back gate voltages $V_{bg}$. Figure 1d, e displays the gate-dependent magneto-transport properties of s6 under perpendicular magnetic field at $T$=2 K. Two sharp transitions at around 3 T and 5 T can be clearly observed on both $R_{xx}$ and $R_{yx}$ in Fig. 1d, e. These two transitions may mark the beginning and ending of the spin-flipping process. With further applying perpendicular magnetic field, the sample is supposed to enter the perfectly aligned FM state [19].

The well quantized Hall resistance plateau with height of 0.99 $h/2e^2$ is detected at -15 T by applying a $V_{bg}$=-17 V, accompanying with the longitudinal resistance as small as 0.004 $h/2e^2$ as shown in Fig. 1d, e. The quantized Hall resistance plateau almost does not change when further increasing $V_{bg}$ to -58 V (within the tolerance of the substrate), which can be clearly observed in Fig. 1f. Besides, the Hall resistance plateau deviates from the quantized value when $V_{bg}$ is above -5 V. The well quantized Hall resistance plateau and nearly vanishing longitudinal resistance are characteristics of high-Chern-number QHE without LLs contributed by dissipationless chiral edge states and indicate a well-defined Chern insulator state with $C$=2, which has never been reported before.

In the absence of magnetic field, MnBi$_2$Te$_4$ bulk is an AFM TI, whose side surfaces are gapless and (111) surfaces are intrinsically gapped by exchange interactions [11, 12, 21]. The gapped surface states are characterized by a quantized Berry phase of $\pi$ and can display the novel half-quantum Hall effect [23, 24]. Due to the AFM nature of the bulk, Hall conductance or topological Chern number of MnBi$_2$Te$_4$ (111) films is dictated by the surface states, which depends critically on the film thickness. For even- and odd-layer films, the two surfaces (on the top and bottom) display half-integer Hall conductance of opposite and identical signs, leading to $C = 0$ and 1, respectively [11]. Obviously, one would never obtain high Chern number $C > 1$ in AFM MnBi$_2$Te$_4$.

However, when MnBi$_2$Te$_4$ is driven from AFM to FM states by external magnetic field, physical properties of the material change dramatically. While the interlayer coupling is restricted by the *PT* (combination of inversion and time-reversal) symmetry in AFM MnBi$_2$Te$_4$ [11, 21], it gets greatly enhanced in the FM state by *PT* symmetry breaking, which generates more dispersive bands along the Γ-Z direction than the AFM state (Fig. S9). Remarkably, the magnetic transition results in a topological phase transition from an AFM TI to a ferromagnetic Weyl semimetal in the bulk [11, 12], leading to a physical scenario to design Chern insulators with $C > 1$ [21, 25-27]. Figure 1g shows the schematic FM order and electronic structure of the $C$=2 Chern insulator state with two chiral edge states across the band gap.

Figure 2 shows the temperature evolution of the high-Chern-number QHE without LLs with the $V_{bg}$=-19 V. As the temperature increases to 13 K, the height of Hall resistance plateau stays above 0.97 $h/2e^2$ and $R_{xx}$ remains below 0.026 $h/2e^2$. With the temperature further increasing to 15 K, the value of Hall resistance plateau reduces to 0.964 $h/2e^2$ and $R_{xx}$ increases to 0.032 $h/2e^2$. This working temperature of the high-Chern-number QHE without LLs is much higher than liquid helium temperature, which shows potential application of QHE in low-dissipation electronics. Furthermore, the High-Chern-number QHE without LLs has also been detected in two more 9-SL devices (Fig. S2-4).

**High-temperature QHE without LLs**

We further study the 7-SL and 8-SL MnBi$_2$Te$_4$ devices (s2 and s3) and the results are displayed in Fig. 3 and Fig. S5. As shown in Fig. S5a, $R_{yx}$ of s2 reaches a well quantized Hall resistance plateau with height of 0.98 $h/e^2$ by applying a small $V_{bg}$=6.5 V at $T$=1.9 K, accompanying with $R_{xx}$ as low as 0.012 $h/e^2$, which is a hallmark of Chern insulator state with $C$=1. When further increasing $V_{bg}$ to 10 V, the quantized Hall resistance plateaus remain robust as shown in Fig. S5a, c. Temperature evolution of $R_{yx}$ and $R_{xx}$ in s2 with $V_{bg}$=6.5 V are shown in Fig. 3a, b. Impressively, as temperature increases, the values of Hall resistance plateau shrink slowly and the plateau can survive up to 45 K (Hall resistance plateau with height of 0.904 $h/e^2$), much higher than the Néel temperature $T_N \sim 21$ K of s2 (Fig. S5b). The high-temperature QHE without LLs is also observed in the 8-SL device s3. As shown in Fig. 3d, e, $R_{yx}$ of s3 is 0.997 $h/e^2$ at 1.9 K ($R_{xx} \sim 0.00006$ $h/e^2$), 8 V, and even at 30 K (above Néel temperature $T_N$=22.5 K), $R_{yx}$ can reach 0.967 $h/e^2$ ($R_{xx} \sim 0.0023$ $h/e^2$). The quantized plateaus from 1.9 K to 30 K are very clear and overlapped. Figure 3c and f displays the color plot of $R_{yx}$ in s2 and s3 as a function of the temperature and magnetic field at $V_{bg}$=6.5 V and 8 V, respectively. Based on the experimental data, the *B-T* phase diagram can be summarized. The phase diagram is characterized by the phase boundaries, $B_{AFM}$ (*T*) and $B_{QH}$ (*T*). The $B_{AFM}$ (*T*) data points, as the boundary of the AFM states, are composed of the peak values of the $R_{xx}$ (*B*) curves (Fig. 3b, e) at various temperatures (the cyan spheres) and the peak value of the $R_{xx}$ (*T*) curve (Fig. S5b and Fig. S8) at zero magnetic field (the pink sphere). The $B_{QH}$ (*T*) curves, as the boundaries of the Chern insulator states (the yellow spheres), represent the magnetic fields required to reach the 99% of the Hall resistance plateau at different temperatures, above which the device is driven to FM state and becomes a Chern insulator with *C*=1. It is obvious that the AFM state disappears at $T_N$. However, the Hall plateau shows nearly quantized resistance even at 45 K (0.904 $h/e^2$) in s2 and 30 K (0.967 $h/e^2$) in s3, which reveals that the Chern insulator state exists at the temperature much higher than $T_N$, indicating a potential way to realize QHE without LLs above liquid nitrogen temperature.

**DISCUSSION**

A fundamental question is that whether the observed quantized Hall resistance plateau is caused by Landau level quantization, as the ordinary QHE with LLs can also give rise to quantized Hall resistance plateau and vanishing $R_{xx}$. We estimate the mobility values of our devices according to the slope of Hall resistance near zero magnetic field [18]. The mobility values are ranging from 100-300 cm$^2$ V$^{-1}$ s$^{-1}$, which are typically below the critical value for formation of LLs at 15 T [28]. To further exclude the possibility of QHE with LLs, we performed controlled measurements by changing the carrier type. In general, the Chern number in ordinary QHE corresponds to the occupancy of LLs and the sign of Chern number will change once the carrier type is switching. However, as shown in Fig. S6c, d, the carrier type in the device s4 (7-SL) with *C*=1 is tuned from *p* to *n* when increasing the back gate voltage from 0 V to 99.5 V, while the sign of Chern number does not change. Furthermore, for the *C*=2 devices, the quantized $R_{yx}$ plateau in device s6 with *n*-type carriers (Fig. 1d) and s7 with *p*-type carriers (Fig. S4a) have the same sign. These observations unambiguously demonstrate that the observed quantized Hall resistance plateau has nothing to do with LLs and the quantized $R_{yx}$ originates from Chern insulator state.

FM MnBi$_2$Te$_4$ belongs to magnetic Weyl semimetals, and has one simple pair of Weyl points (WPs) along the Γ-Z direction located at $k_W$ and $-k_W$. The $k_z$-dependent

Chern number $C(k_z) = \sigma_{xy}(k_z) h/e^2$ defined for 2D momentum planes with specified $k_z$ must be quantized (except at the gapless WPs) and abruptly jumps at positions of Weyl points. $C(k_z)$ equals one between the two Weyl points due to topological band inversion and zero elsewhere as illustrated in Fig. 4a. The averaged anomalous Hall conductance per unit layer is given by

$$\bar{\sigma}_{xy} = \frac{c_0}{2\pi} \int_{-c_0/\pi}^{c_0/\pi} \sigma_{xy}(k_z) dk_z = |\tilde{k}_W| e^2/h,$$

where $c_0$ is the out-of-plane thickness of each SL, and $\tilde{k}_W = |k_W| c_0/\pi$. For a $N$-layer FM thin film, its electronic states can be viewed as quantum-well states and possess a finite band gap due to quantum confinement. Generally, $\sigma_{xy}$ of thin films would grow with film thickness, as its ideal bulk contribution is $N|\tilde{k}_W|e^2/h$. On the other hand, $\sigma_{xy}$ of 2D gapped films must take quantized values $C(N)e^2/h$ as topologically required. Therefore, for thick films with minor surface effects, the thickness-dependent Chern number $C(N)$ would change discretely by 1 for every $\Delta N = 1/|\tilde{k}_W|$, implying that high Chern number is feasible by increasing film thickness. The discrete increase of Chern number with increasing film thickness is a generic feature of ferromagnetic Weyl semimetals, which can also be understood by the topological band inversion picture as discussed in Methods.

The above physical picture is confirmed by the first-principles study, which gives $\tilde{k}_W = 0.256 \approx 1/4$ for the bulk and shows that $C(N)$ indeed increases by 1 for every $\Delta N = 4$ (Fig. 4b). Note that it is theoretically challenging to accurately predict $C(N)$, since the predicted $\tilde{k}_W$ depends sensitively on the exchange-correlational functional and the lattice structure. Based on the modified Becke-Johnson (mBJ) functional [29], we systematically tested the influence of lattice parameter $c_0$ on band structure and $C(N)$ (Fig. S9), and finally decided to use the experimental value $c_0 = 13.6$ Å. As shown in Fig. 4b, the 9-SL film is a high-Chern-number band insulator with $C = 2$. Compared to the AFM films studied before [11], band structure of the FM film displays much more pronounced quantum confinement effects, as visualized by significant band splitting between quantum well states (Fig. 4c). A quantum confinement induced gap ~5 meV is located at the $\Gamma$ point. The edge-state calculation reveals that there exist two chiral gapless edge channels within the gap (Fig. 4d), which confirms $C = 2$. Therefore, first-principles calculations indicate that high-Chern-number band insulators can be realized in the FM Weyl semimetal $MnBi_2Te_4$ by means of quantum confinement.

The theory suggests that topological Chern number is tunable by controlling film thickness of FM $MnBi_2Te_4$. By reducing the film thickness to 7-SL, Chern number decreases to $C = 1$, as found experimentally. 8-SL is the marginal case, which has $C = 1$ in experiment and $C = 2$ in theory. The discrepancy is possibly caused by the surface/interface effects that are not theoretically considered. Contrariwise, the increase of film thickness could lead to higher Chern numbers ($C > 2$), which is awaiting experimental confirmation. Moreover, since the Chern insulator phase appears in the FM state, the weak inter-SL antiferromagnetic exchange coupling is irrelevant to the topological physics. Thus, the working temperature of QHE without LLs will not be limited by the Néel temperature, and can be quite high due to the strong, ordered magnetism of $MnBi_2Te_4$.

**CONCLUSIONS**

In summary, we discovered high-Chern-number QHE ($C$=2) without LLs showing two sets of dissipationless chiral edge states above 10 K and $C$=1 Chern insulator state above the Néel temperature, which is also the highest temperature for QHE without LLs. Our findings open a new path for exploring the interaction between topology and magnetism, as well as the potential application of topological quantum states in low-power-consumption electronics at higher temperatures.

**METHODS**

**Crystal growth**

High-quality $MnBi_2Te_4$ single crystals were grown by directly reacting a stoichiometric mixture of high-purity $Bi_2Te_3$ and MnTe, which were prepared by reacting high-purity Bi (99.99%, Adamas) and Te (99.999%, Aladdin), and Mn (99.95%, Alfa Aesar) and Te (99.999%, Aladdin), respectively. The reactants were sealed in a silica ampoule under a dynamic vacuum, which was then heated to 973 K and slowly cooled down to 864 K, followed by the prolonged annealing at the same temperature over a month. The quality of mm-sized $MnBi_2Te_4$ crystals was examined on a PANalytical Empyrean diffractometer with Cu Kα radiation.

**Devices fabrication**

The $MnBi_2Te_4$ nanoflakes on 300 nm $SiO_2$/Si substrate were mechanically exfoliated from high quality single crystals using scotch tape. The substrates were pre-cleaned in oxygen plasma for 5 minutes with ~60 mtorr pressure. To obtain flakes with thickness down to several nanometers, we heated the substrate after covering the scotch tape at 393 K (120 ℃) for 1 minute. Standard electron beam lithography in a FEI Helios NanoLab 600i Dual Beam System was used to define electrodes after spin-coating PMMA resist. Then, metal electrodes (Ti/Au or Cr/Au, 65/180 nm) were deposited in a LJUHV E-400L E-Beam Evaporator after Ar plasma cleaning.

**Transport measurements**

Electrical transport measurements were conducted in a 16T-Physical Property Measurement System (PPMS-16T) from Quantum Design with base temperature $T$=1.9 K and magnetic field up to 16 T. Stanford Research Systems SR830 lock-in amplifiers were used to measure longitudinal resistance and Hall signals of the device with an AC bias current of 100 nA at a frequency of 3.777 Hz. The back gate voltages were applied by a Kethiley 2912A source meter.

**First-principles calculations**

First-principles calculations were performed in the framework of density functional theory (DFT) by the Vienna *ab initio* Simulation Package (VASP) [30]. The plane-wave basis with an energy cutoff of 350 eV, the projector augmented wave method together with the Monkhorst-Pack $k$-point mesh of 9×9×5 were used. Typically, the DFT+$U$ method was applied in previous studies of $MnBi_2Te_4$ [11, 21]. Here, to improve the description of electronic band structure, the modified Becke-Johnson (mBJ) functional [29] was employed to study ferromagnetic bulk $MnBi_2Te_4$. Since the mBJ functional cannot be directly applied to describe systems with vacuum space, the tight-binding method was used to model thin films. Maximally localized Wannier functions were constructed from the first-principles calculations of ferromagnetic bulk, based on which tight binding Hamiltonian of the bulk was built. Then, tight binding Hamiltonians of thin films were obtained by means of cutting slabs from the bulk. Specifically, the Hamiltonian of a slab was directly extracted from that of the periodic bulk by setting the coupling between the slab and its neighboring bulk to zero. The minor influence of the surface was neglected during the process. The tight-binding method for thin films was systematically tested and proved to be able to well reproduce DFT results of variant exchange-correlation functionals for different van der Waals materials (e.g. $MnBi_2Te_4$ and $Bi_2Te_3$), provided that the film is thick enough (roughly >3 nm) for safely neglecting the surface effects. Based on the tight-binding Hamiltonian, Chern numbers, band structures and topological edge states were computed by using the WannierTools package [31]. Since the position of Weyl points in momentum space and the topological Chern number of thin films depend sensitively on the out-of-lattice constant $c = 3c_0$, structures with different $c_0$ varying from the theoretical ($c_0 = 13.53$ Å) [11] to experimental ($c_0 = 13.6$ Å) [32] values were systematically studied and compared (Fig. S9).

**ACKNOWLEDGEMENTS**

We thank Pu Yang and Zeyan Yang for help in devices fabrications, and Jiawei Luo and Jiawei Zhang for helpful discussion in transport measurements.

**FUNDING**

This work was financially supported by the National Key R&D Program of China (2018YFA0305600, 2017YFA0303300, and 2018YFA0307100), the National Natural Science Foundation of China (Grant No. 11888101, Grant No. 11774008, Grant No. 51788104, Grant No. 11874035, Grant No. 21975140, and U1832218), and Beijing Natural Science Foundation (Z180010), the Strategic Priority Research Program of Chinese Academy of Sciences (XDB28000000).

**Author contributions**

J. W. conceived and supervised the experiments. J. G. and Y. L. performed transport measurements. Y. X. and J. L. carried out theoretical calculations. H. L. and Y. W. grew the MnBi$_2$Te$_4$ bulk crystals. J. G. fabricated devices. J. G., Y. L., T. L. and J. W. analyzed the data. J. G., Y. L., J. L., Y. X. and J. W. wrote the manuscript with input from all authors.

**Data Availability Statement**

All data analyzed to evaluate the conclusions are available from the authors upon reasonable request.

**Additional Information**

The authors declare no competing financial interests. Correspondence and requests for materials should be addressed to J.W. (jianwangphysics@pku.edu.cn) and Y. X. (yongxu@mail.tsinghua.edu.cn).

**Figure captions**

**Figure 1.** Gate-dependent transport properties of the 10-SL MnBi$_2$Te$_4$ device s6. (a) Crystal structure of MnBi$_2$Te$_4$. The red and blue arrows denote magnetic moment directions of Mn ions. (b) Optical image of the 10-SL MnBi$_2$Te$_4$ device s6. Scale bar represents 10 μm. (c) Temperature dependence of $R_{xx}$ at $V_{bg}$=0 V. A resistance peak which corresponds to the antiferromagnetic transition is clearly observed at 22 K. (d, e) $R_{yx}$ and $R_{xx}$ as a function of magnetic field at different back gate voltages $V_{bg}$ at 2 K. Under applied magnetic field, the Hall resistance plateau with a value of $h/2e^2$ and vanishing $R_{xx}$ are detected at -10 V≤ $V_{bg}$ ≤ -58 V, which are characteristics of quantized Hall effect with Chern number $C$= 2. The black and red traces represent magnetic field sweeping to the positive and negative directions, respectively. (f) $R_{xx}$ and $R_{yx}$ as a function of $V_{bg}$ at 2 K and -15 T. (g) The schematic FM order and electronic structure of the $C$=2 Chern insulator state with two chiral edge states across the band gap. The grey and blue colors are used to distinguish the adjacent MnBi$_2$Te$_4$ SLs.

**Figure 2.** Temperature dependence of the high-Chern-number QHE without LLs in s6. (a, b) $R_{yx}$ and $R_{xx}$ as a function of magnetic field at different temperatures from 2 K to 15 K. The height of Hall resistance plateau can reach 0.97 $h/2e^2$ at 13 K.

**Figure 3.** High-temperature QHE without LLs in MnBi$_2$Te$_4$ devices s2 (7-SL) and s3 (8-SL). (a, b) Temperature dependence of the $C$=1 QHE without LLs in s2 at $V_{bg}$ =6.5 V. The nearly quantized Hall resistance plateau can stay at the temperature up to 45 K

(Hall resistance plateau of 0.904 $h/e^2$). (c) *B-T* phase diagram of s2. The AFM state disappears at $T_N$ ~ 21 K and the *C*=1 QHE state can survive up to 45 K (Hall resistance plateau of 0.904 $h/e^2$), much higher than $T_N$. (d, e) $R_{yx}$ and $R_{xx}$ as a function of magnetic field in s3 at various temperatures at $V_{bg}$ =8 V. The well-defined quantized Hall resistance plateau can stay at the temperature as high as 30 K (Hall resistance plateau of 0.967 $h/e^2$). (f) *B-T* phase diagram of s3. The AFM state disappears at $T_N$ ~ 22.5 K and the well-defined quantization can stay till 30 K (Hall resistance plateau of 0.967 $h/e^2$).

**Figure 4.** Theoretical calculations of 9-SL FM MnBi$_2$Te$_4$. (a) The illustration of band structure along the $k_z$ direction and the $k_z$-dependent Chern number in the FM bulk phase of MnBi$_2$Te$_4$, which is a magnetic Weyl semimetal. Weyl points (WPs) with topological charge of +1 and -1 are denoted by blue and red circles in the top panel, respectively. The positions of WPs correspond to the jump of Chern number in the bottom panel. (b) Chern number as a function of film thickness. (c, d) Band structure and edge states along the (100) direction in the 9-SL film.

Figures

Figure 1

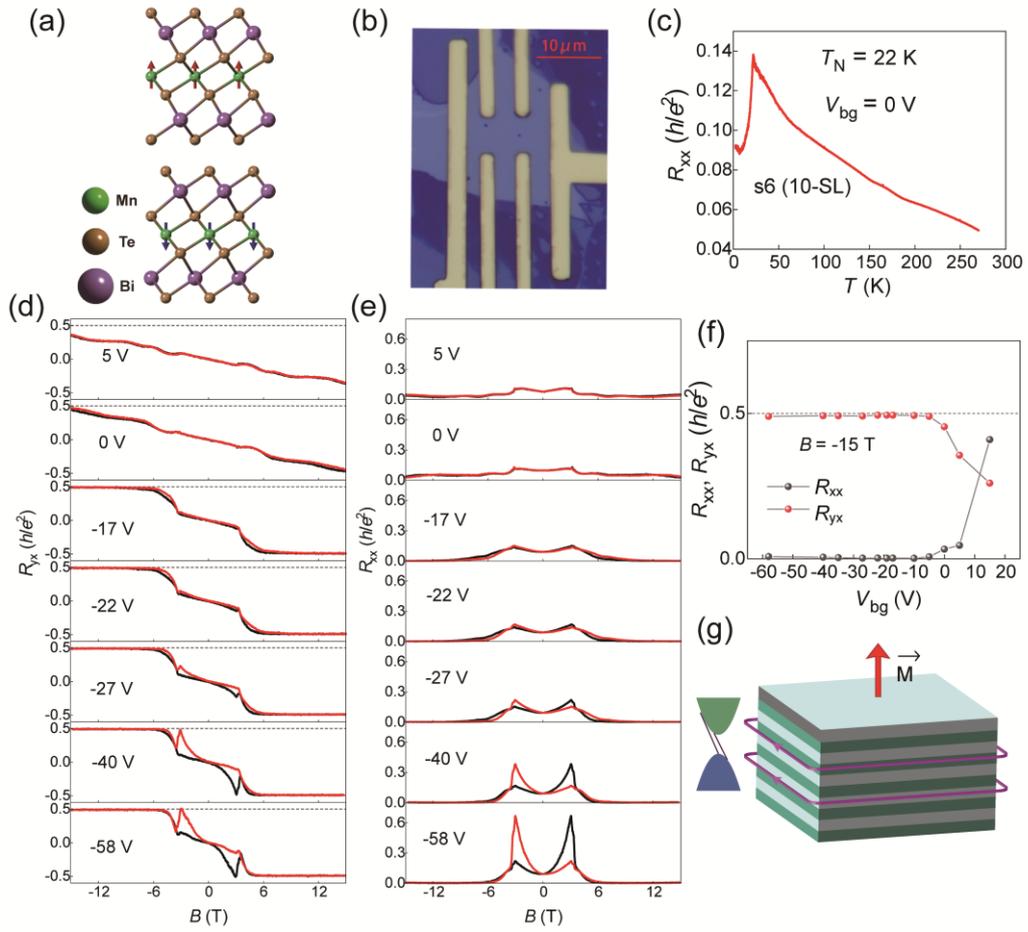

Figure 2

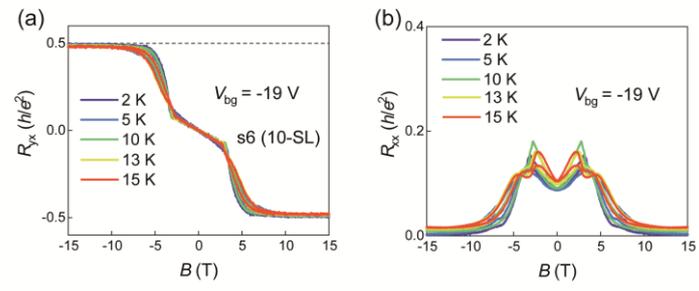

Figure 3

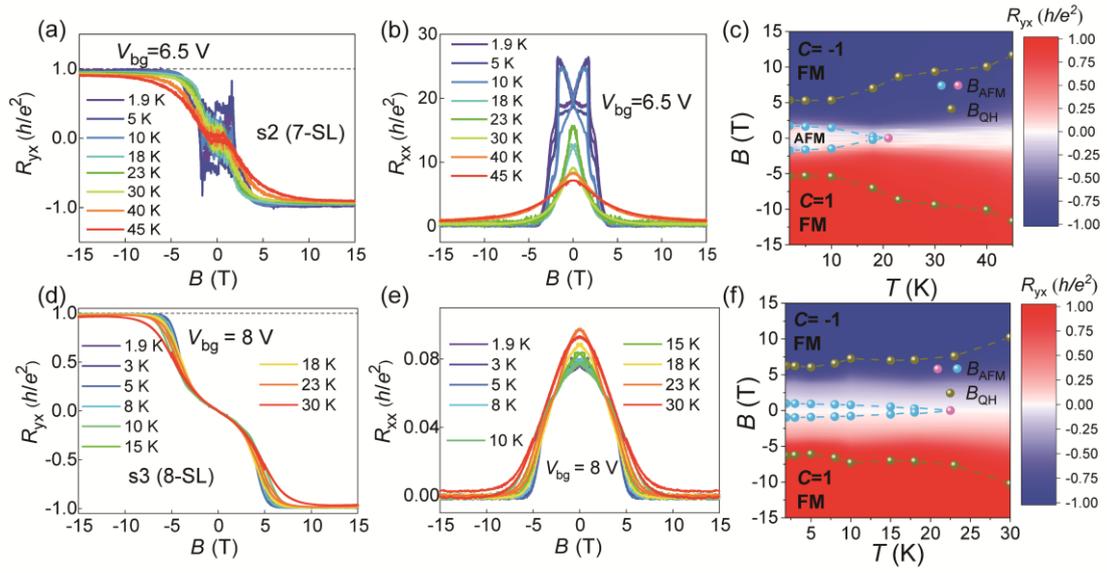

Figure 4

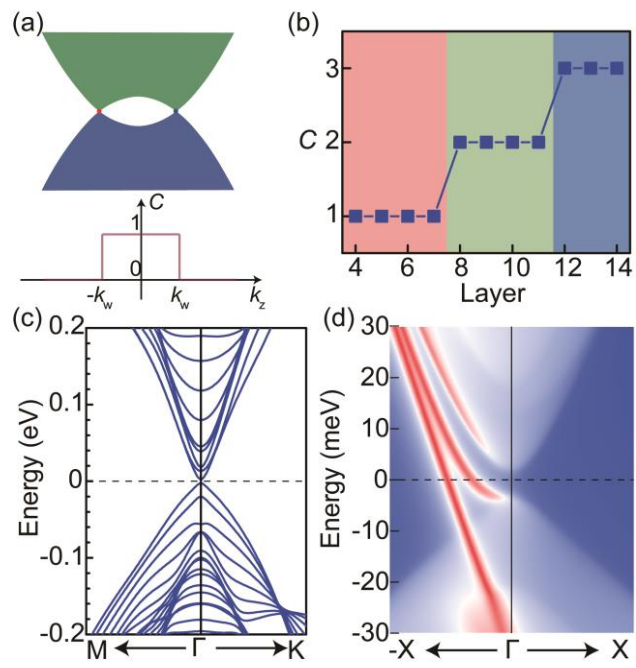